\documentstyle[11pt,newpasp,twoside,psfig]{article}
\index{summary}
\index{instructions}
\index{template}

\def\edcomment#1{\iffalse\marginpar{\raggedright\sl#1\/}\else\relax\fi}
\reversemarginpar

\newcommand{\chandra}{{\em Chandra}}
\newcommand{\rosat}{{\em ROSAT}}
\newcommand{\asca}{{\em ASCA}}

\newcommand{\etal}{\mbox{{\em et~al.}}}
\newcommand{\NH}{\mbox {$N_{\rm H}$}}

\begin{document}
\title{CHANDRA OBSERVATIONS OF THE EASTERN LIMB OF THE  VELA SUPERNOVA
REMNANT}
\author{P.P.~Plucinsky, R.K~Smith, R.J.~Edgar, T.J.~Gaetz, P.O.~Slane}
\affil{Harvard-Smithsonian Center for Astrophysics, 60 Garden St.,
MS-70, Cambridge, MA 02138}
\author{W.P.~Blair}
\affil{Department of Physics and Astronomy, Johns Hopkins University,
3400 N. Charles St., Baltimore, MD 21218-2686}
\author{L.K.~Townsley, and P.S.~Broos}
\affil{Department of Astronomy \& Astrophysics, The Pennsylvania 
State University, 525 Davey Lab, University Park, PA 16802}

\begin{abstract}
  We present results from two \chandra/ACIS observations of the
so-called Vela ``Bullet D'' region on the eastern limb of the Vela
supernova remnant.  The Bullet D region is a bright X-ray feature,
identified by Aschenbach et al. (1995) from the ROSAT All-Sky Survey,
which protrudes beyond the blast wave on the eastern side of the
remnant.  It has been suggested that this feature is a fragment of
supernova ejecta which is just now pushing beyond the position of
the main blast wave.
An alternate explanation is that the feature is a ``break-out'' of the
shock in which inhomogeneities in the ambient medium cause the shock
to be non-spherical.  The \chandra\/ image shows a fragmented,
filamentary morphology within this region.  The \chandra\/ spectra
show strong emission lines of O, Ne, and Mg.  
Equilibrium ionization models indicate
that the O and Ne abundances are significantly enhanced compared
to solar values.  However, non-equilibrium ionization models can fit
the data with solar O abundances and Ne abundances enhanced by only a
factor of two.  The \chandra\/ data are more consistent with the
shock breakout hypothesis, although they cannot exclude the fragment
of ejecta hypothesis.

%, and also with enhanced abundances.  

%  Based on these spectral results and the morphology, we
%conclude that the Bullet D feature is more likely to be the result of
%the shock interacting with an inhomogeneous medium.

\end{abstract}

\vspace{-0.25in}

\section{Introduction}
%\vspace{-0.15in}

%- What is Vela
%- Distance to Vela
%- Aschenbach's discovery and interpretation
%- Tsunemi and Miyata studies
%- Bullet D is the brightest

The Vela supernova remnant (SNR) is a large (diameter $\sim 8$ degrees), nearby
supernova remnant associated with the Vela pulsar and is one of the
brightest objects in the X-ray sky.
Recent work indicates the distance to Vela is only $250\pm30$~pc
(Cha~\etal\/ 1999 and Jenkins \& Wallerstein 1995), making it an ideal
candidate for resolving fine structure in X-rays with the {\em Chandra
X-ray Observatory}.
\rosat\/ All-sky Survey observations showed a
complicated morphology and revealed the outer extent of the remnant
for the first time.
Aschenbach~\etal\/~(1995) identified several features protruding
beyond what is believed to be the primary blastwave as ``explosion
fragments''.
The brightest of these features is the so-called ``Bullet D''.  
Tsunemi~\etal\/~1999 and Miyata~\etal\/~2001 observed ``Bullet A''
with \asca\/ and \chandra\/ respectively and detected strong Si 
emission lines which they concluded was evidence that Bullet~A
was indeed composed of ejecta from the original explosion.
Moriguchi~\etal (2001) suggested that the bullet features were more
likely the result of the interaction of the SNR shock with an
inhomogeneous medium as indicated by the numerous molecular clouds
identified by a CO survey of the region.
% Redman~\etal\/ 2000 claim that the optical data suggest an
%association between G266.2--1.2 and Bullet D.
Redman~\etal\/ (2000) noted the coincidence of the bright optical
filamentary nebula RCW~37 and Bullet~D.
We proposed two
\chandra\/ observations, one  at the head of the bullet and the other
in the ``wake'' in order to examine the proposed explanations for
Bullet~D.

\vspace{-0.15in}
\section{Observations}
%\vspace{-0.15in}

%- two chandra observations, describe in words when they were
%- show image
%- show spectral extraction regions

\chandra\/ observed the eastern limb of Vela in September 2000 with
the ACIS-I detector.  The exposure at the head of the Bullet D was
31~ks and the exposure in the wake region was 52 ks. Figure~1 
displays the image from the head of Bullet D after selecting events
in the 0.4 to 2.5~keV range, binning the data in $2\times2\arcsec$
pixels, and smoothing the data with a Gaussian filter with a FWHM of
3\arcsec.  The data show a fragmented, filamentary structure
consistent with a shock interacting with an inhomogeneous medium.
If this feature were a fragment of ejecta from the explosion, one 
would expect a larger contrast in intensity of the X-ray emission
from the head of the feature to the trailing edges as seen in
Bullet~A. 

\begin{figure}[htb]

\hbox{

\vbox{\psfig{figure=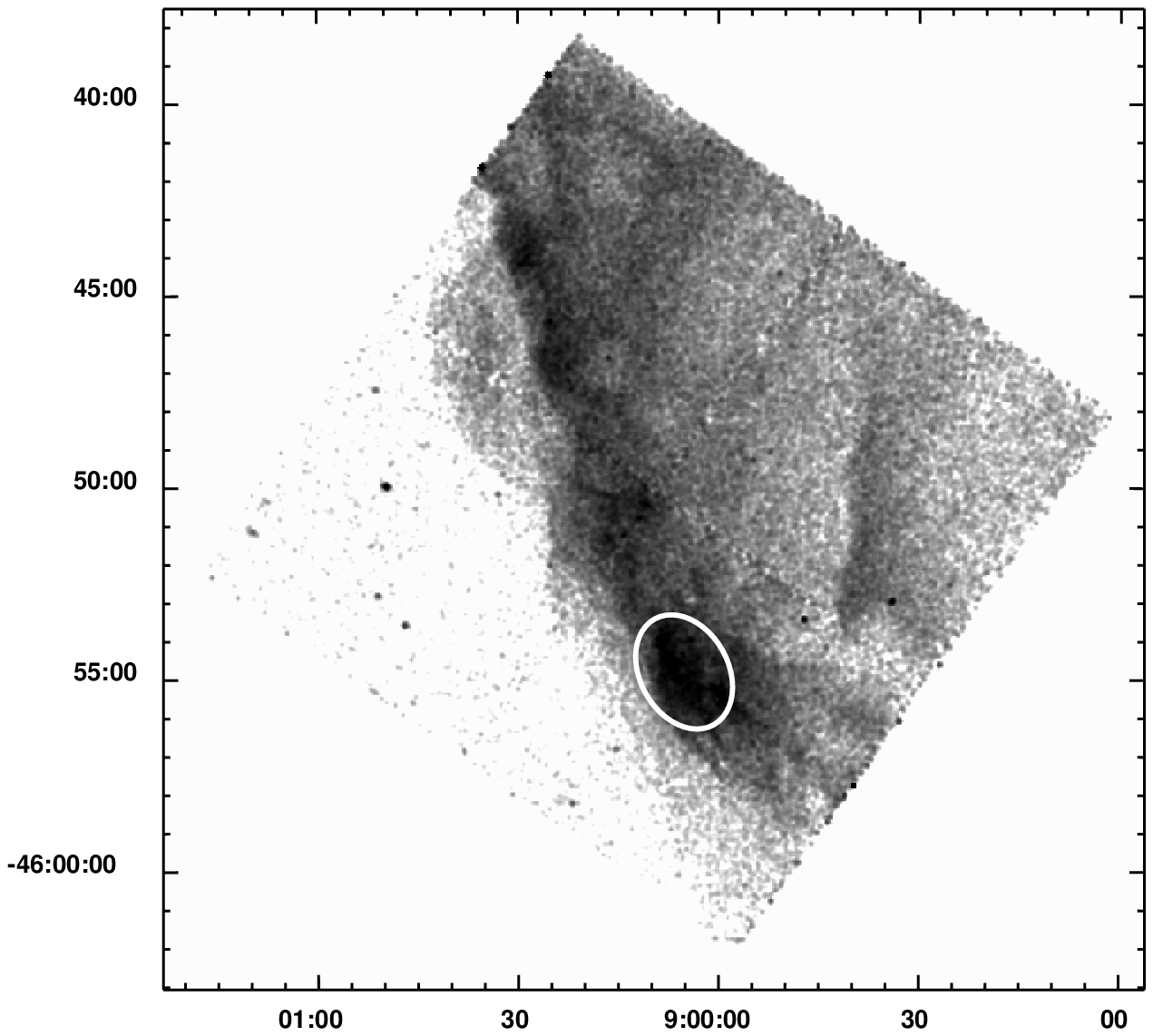,width=2.4in}}

\hfill

\vbox{\psfig{figure=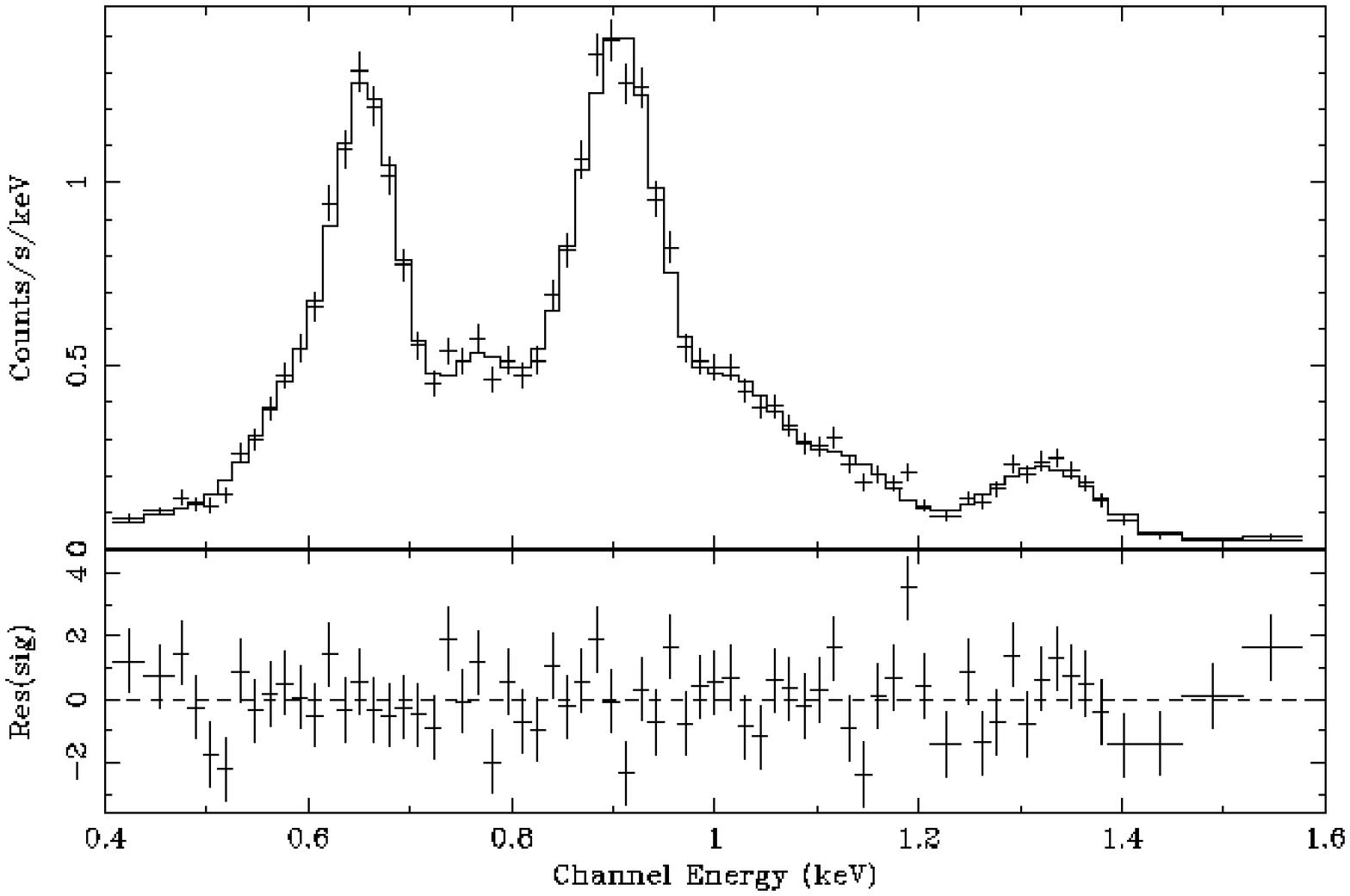,width=2.75in}}

}

\caption{ (LEFT) {\it Chandra} ACIS-I observation of the Eastern Limb of
 Vela.  The data have been filtered to include energies between 0.4
 and 2.5 keV and binned into $2\times2\arcsec$ pixels.  The data
have been smoothed with a Gaussian filter with a FWHM of $3\arcsec$.
The ellipse indicates the spectral extraction region. Figure 2:
 (RIGHT) Spectrum extracted from the ellipse in Figure~1 after
 applying the CTI correction.
}

%\plotfiddle{}{4.5in}{-90.0}{50}{40}{-230.0}{680.0}
\vspace{-0.15in}
\end{figure}

\setcounter{figure}{2}

\vspace{-0.15in}

\section{Spectral Analysis}

%\vspace{-0.15in}
%- spectra preparation, need CTI correction
%- fits with equilibrium models
%- fits with NEI models
%- what is the NH, refer to Dubner
%- what is the kT
%- what are the abundances

The frontside-illuminated (FI) CCDs on ACIS suffered radiation damage
early in the mission which resulted in a large increase in the
charge transfer-inefficiency (CTI) and a significant reduction in the
spectral resolution.  The observations of the Bullet D are seriously
impaired by this damage since the object covers all
four CCDs on the ACIS-I array.  We have used the CTI-correction software
and appropriate response matrices developed at Penn State
(see Townsley~\etal\/ 2000, Townsley~\etal\/ 2000a, and Townsley~\etal\/
2000b).  The CTI correction dramatically improves
the quality of the data as demonstrated in our analysis of the
spectra of the SMC SNR 1E0102.2-7219 (see Plucinsky~\etal\/ 2001).
Fortunately, the spectrum of the Bullet D is quite similar to that
of 1E0102.2-7219 in that both have strong lines of O, Ne, and Mg with
little or no Fe.

% We have extracted spectra from various locations from the Bullet D
%head shown in Figure~1 and from the wake pointing.  We will defer a
%discussion of the comparison of these spectra to a future paper and
We focus on the spectrum of the brightest part of the Bullet D
feature described by the ellipse in Figure~1.  The ACIS spectrum of
this feature is shown in Figure~2.  The dominant features are the
O{\small VIII}~${\rm Ly~\alpha}$ line at 654~eV, the Ne{\small IX} triplet at 
$\sim910$~eV, and the MgXI triplet at $\sim1.34$~keV.  We fit these data
with a collisional ionization equilibrium (CIE) model and derived a best-fit
temperature of $\sim0.45$~keV after fixing the \NH\/ to be
$3.0\times10^{20} {\rm cm^{-2}}$ (see Dubner~\etal\/ 1998).   At this 
temperature, the CIE model
indicates that the O and Ne abundances must be enhanced by several
times over solar values.

\begin{figure}[htb]

\hbox{

\vbox{\psfig{figure=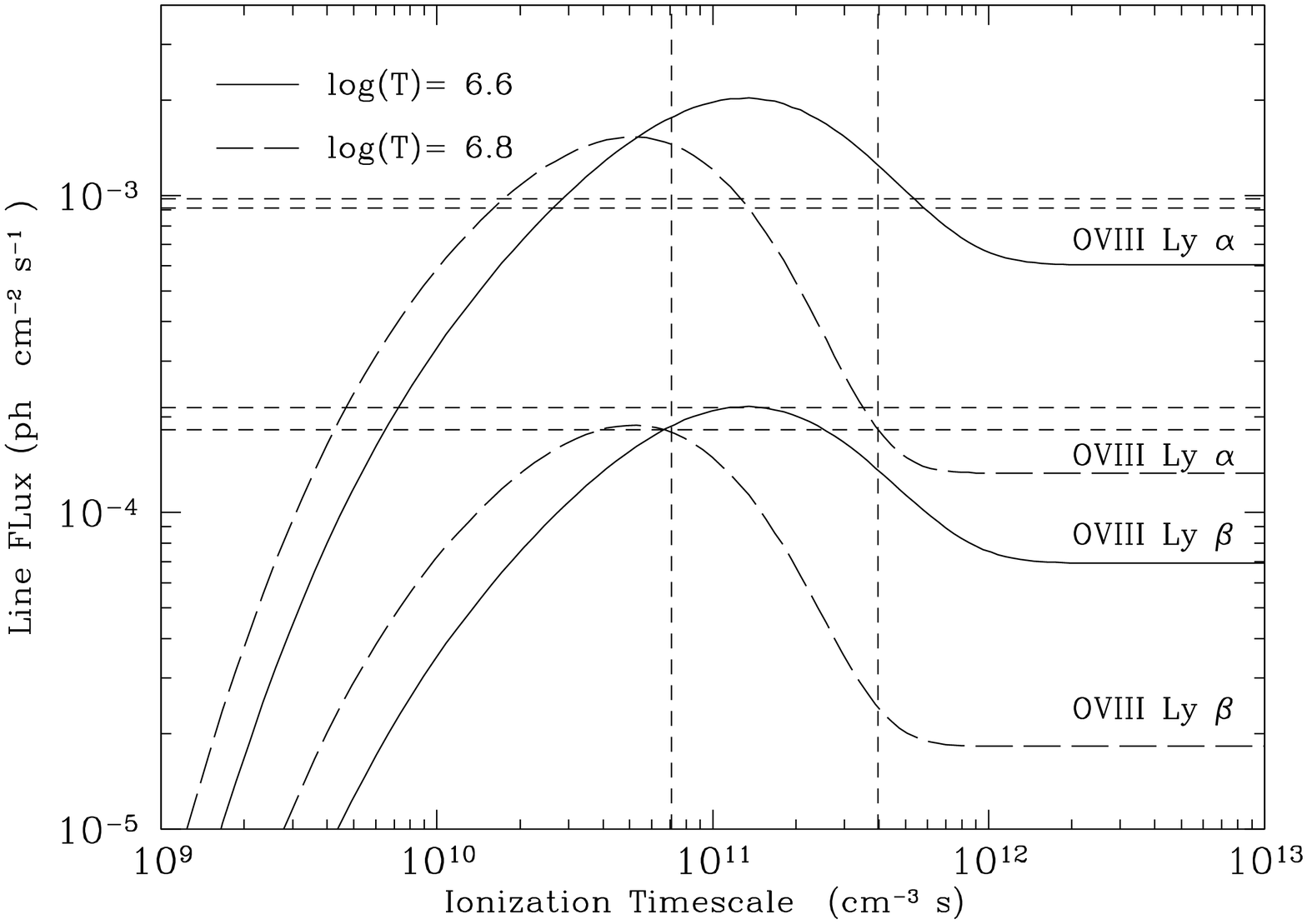,width=2.6in}}

\hfill

\vbox{\psfig{figure=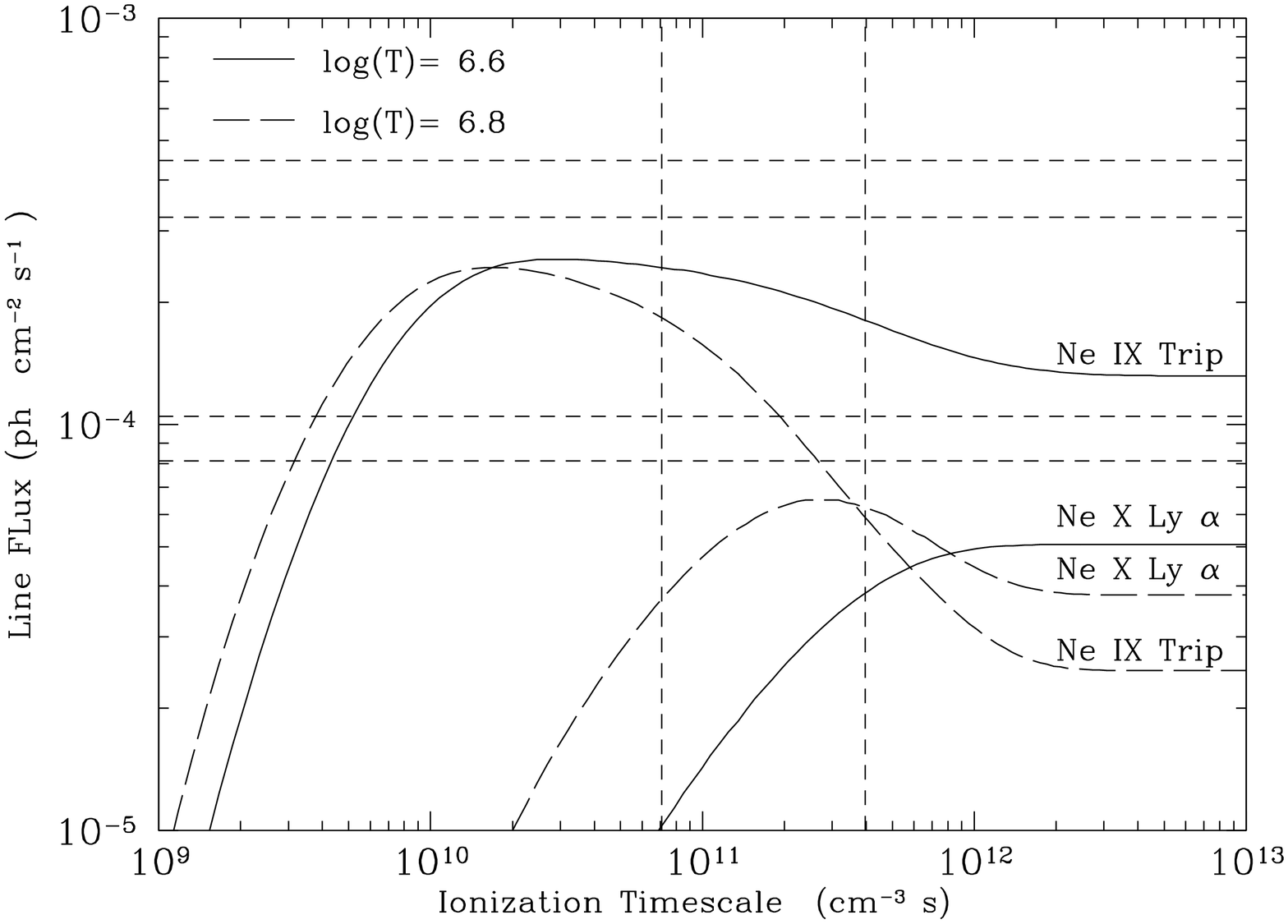,width=2.6in}}

}

\caption{ (LEFT) The predicted and measured emissivities of the 
O{\small VIII}~${\rm Ly~\alpha}$ and ${\rm Ly~\beta}$ lines
as a function of ionization timescale and temperature.
The measured emissivities are indicated by the dashed horizontal
lines and the region of ionization timescales discussed in the
text are indicated by the vertical dashed lines. Figure 4: (RIGHT)
Same as Figure~3 except for the the Ne{\small IX} triplet and 
Ne{\small X}~${\rm Ly~\alpha}$ line. }

%\plotfiddle{}{4.5in}{-90.0}{50}{40}{-230.0}{680.0}
\vspace{-0.10in}
\end{figure}

Next we fit the data with a model of a simple continuum plus Gaussians to
represent the line emission.  We used the ratios of 
O{\small VII}~triplet/O{\small VIII}~${\rm Ly~\alpha}$ and 
Ne{\small IX}~triplet/Ne{\small X}~${\rm
Ly~\alpha}$ as non-equilibrium diagnostics.  Both diagnostics
indicated that temperatures below log(T)=6.4 are not consistent with 
the data.  Both the O and Ne line ratios are consistent with 
ionization timescales in the range  $7.0\times10^{10}< n_et <
7.0\times10^{11}~{\rm cm^{-3} s}$.  We can impose an upper limit on the
ionization timescale  of $n_et=4.0\times10^{11}~{\rm cm^{-3} s}$
by assuming an age of 12,000~yr and a maximum ambient density of 
$n_o=1.0~{\rm cm^{-3}}$.  We can then compare the emissivities of the 
O{\small VIII}~${\rm Ly~\alpha}$, O{\small VIII}~${\rm Ly~\beta}$,
Ne{\small IX}~triplet, and Ne{\small X}~${\rm Ly~\alpha}$ lines to
the expected values.  The measured emissivities are plotted
against the expected emissivities as a function
of temperature and ionization timescale in Figures~3 and~4.
The measured Ne emissivity is higher than the predicted emissivity for
all values of the ionization timescale regardless of temperature,
indicating that the abundance of Ne must be higher than solar.  In
the preferred range of ionization timescales from $7.0\times10^{10}< 
n_et < 4.0\times10^{11}~{\rm cm^{-3} s}$, the Ne abundance
required to be consistent with the model predictions varies
from 1.6 to $6.4~\times$ solar.
The situation with O is significantly
different.  In the range of preferred ionization timescales from 
$7.0\times10^{10}< n_et < 4.0\times10^{11}~{\rm cm^{-3} s}$, the
measured emissivity is lower than the expected emissivity for a large
range of ionization timescales.
Over this range, the O abundance required to be consistent with the
model predictions varies from 0.5 to $5.0~\times$ solar.
The O and Ne data are most consistent with each other for a
temperature of log(T) = 6.6 and an ionization timescale of
$4.0\times10^{11}~{\rm cm^{-3} s}$.  At these values the Ne abundance is
$\sim$ twice solar and the O abundance is $\sim$ solar.
%These results indicate that the Chandra spectra are
%consistent with a solar or even sub-solar abundance of O.

\section{Discussion}
%\vspace{-0.15in}
The morphology of the X-ray emission revealed by
\chandra\/ is more consistent with a 
SNR shock interacting with an inhomogeneous medium than with a
discrete fragment of ejecta.  The \chandra\/ spectra indicate that
Ne is enhanced above solar values, but at a rather modest level,
and O is consistent with the solar value.  
We therefore conclude the \chandra\/ observations of the Vela Bullet D
region are more
consistent with a shock breakout hypothesis than with a bullet of
ejecta hypothesis, although the \chandra\/ data cannot rule out the
bullet idea. 

\vspace{0.10in}
\acknowledgments 
%\vspace{-0.15in}

We thank
all of the engineers, technicians, and scientists who have made the
{\em Chandra X-ray Observatory} such a success. PPP, RKS, RJE, TJG, and POS 
acknowledge support for this work from NASA contracts NAS8-39703 and
GO0-1127.
LKT and PSB acknowledge support for this work from NASA contract
NAS8-38252.

%- is this a fragment from the original explosion
%- Optical/UV indicate a normal shock, no enhanced abundances
%- Pressures are quite different
%- another SNR ??
%- no relationship to G266.2

\end{document}